\documentclass[aps,pre,twocolumn,eqsecnum,superscriptaddress]{revtex4}    
\usepackage{amsmath}    
\usepackage{amssymb}
\usepackage{graphicx}   
\usepackage{subfigure}

\begin{document}

\title{Sacrificial bonds and hidden length in biomaterials -- a kinetic, constitutive description
of strength and toughness in bone}
\author{Charles K. C. Lieou}
\affiliation{Department of Physics, University of California, Santa Barbara, CA 93106, USA}
\author{Ahmed E. Elbanna}
\affiliation{Department of Physics, University of California, Santa Barbara, CA 93106, USA}
\affiliation{Department of Civil and Environmental Engineering, University of Illinois, Urbana-Champaign, IL 61801, USA}
\author{Jean M. Carlson}
\affiliation{Department of Physics, University of California, Santa Barbara, CA 93106, USA}
\date{\today}

\begin{abstract}
Sacrificial bonds and hidden length in structural molecules account for the greatly increased fracture toughness of biological materials compared to synthetic materials without such structural features, by providing a molecular-scale mechanism for energy dissipation. One example is in the polymeric glue connection between collagen fibrils in animal bone. In this paper, we propose a simple kinetic model that describes the breakage of sacrificial bonds and the release of hidden length, based on Bell's theory. We postulate a master equation governing the rates of bond breakage and formation. This enables us to predict the mechanical behavior of a quasi-one-dimensional ensemble of polymers at different stretching rates. We find that both the rupture peak heights and maximum stretching distance increase with the stretching rate. In addition, our theory naturally permits the possibility of self-healing in such biological structures.
\end{abstract}

\maketitle

\section{Introduction}

Many biological, polymeric materials gain their strength and toughness through the formation of sacrificial bonds and hidden length. Examples include bone~\cite{smith_1999,fantner_2005,hansma_2005,fantner_2006,thompson_2001,launey_2010,buehler_2010}, abalone shells~\cite{smith_1999,currey_1977,jackson_1988} and diatoms~\cite{gebeshuber_2003,higgins_2002,dudgale_2005,sarkar_2007}. Often, sacrificial bonds connect two different sites on a molecular backbone, thereby constraining part of the polymer from stretching. These bonds are typically weaker than the covalent bonds on the molecular backbone; they break and release ``hidden length'' before the molecular backbone ruptures.  This molecular-scale mechanism has been found to greatly increase the total amount of work needed to break the material.

\begin{figure}[here]
\centering \includegraphics[width=.40\textwidth]{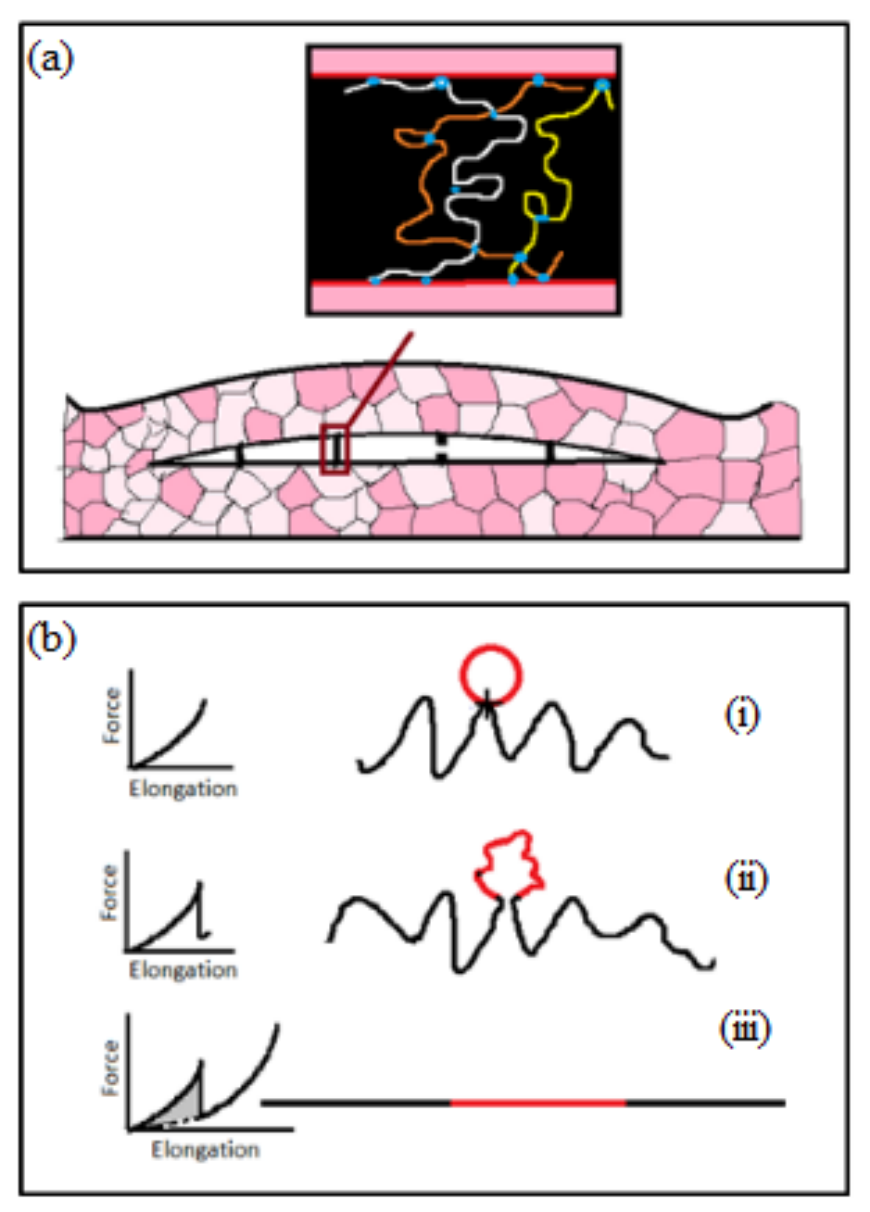} \caption{\label{fig:sbhl}(Color online) Basic features of sacrificial bonds and hidden length. (a) Hypothesized structure of polymeric glue strands, with sacrificial bonds (blue circles) along the polymer chain backbone resisting chain rupture as they are stretched during microcracking. There are three types of sacrificial bonds: bonds within a polymer chain; bonds between different chains; and bonds connecting the polymer chain to the substrate (collagen fibrils in the case of animal bone). Adapted from~\cite{fantner_2005}. (b) Force change associated with sacrificial bond breakage and hidden length release. (i) Before a sacrificial bond is broken, only the black length of the molecule contributes to the entropic configurations and to the force with which the molecule resists stretching. The red length of the molecule is hidden from the force by the sacrificial bond. (ii) When the bond breakage threshold is reached, the bond breaks and the whole length of the polymer (black plus red) contributes to its configurational entropy. This sudden increase in entropy leads to an abrupt force drop. (iii) As the polymer molecule is further stretched, the force it supports increases, until the entire molecule detaches from the substrate and ruptures. The grey area represents the extra work done in stretching a polymer with sacrificial bonds and hidden lengths, relative to a polymer of the same length but without such structural features. Reprinted with permission from~\cite{elbanna_2012}.}
\end{figure}

An important example of sacrificial bonds and hidden length occurs in the polymeric glue connection between collagen fibrils in animal bone~\cite{smith_1999,fantner_2005,hansma_2005,fantner_2006,thompson_2001,launey_2010,buehler_2010}, illustrated schematically in Fig.~\ref{fig:sbhl}. Each intact sacrificial bond shields part of the glue strand from contributing to its end-to-end distance. Given an end-to-end distance, a glue strand of smaller apparent length carries less entropy than one with more available length. In other words, the presence of sacrificial bonds and hidden length amplifies the amount of force that is necessary to stretch the polymers, and therefore accounts for the increase in fracture toughness of the material. Following breakage of one sacrificial bond, the corresponding hidden length now unravels, causing a force drop as an immediate result of the spike in entropy.

We recently proposed a theoretical model that accounts for this mechanism and captures the mechanical response of the stretched polymer network in the quasi-static limit~\cite{elbanna_2012}. In that model, we assume that the strength of sacrificial bonds is a random variable, primarily to account for the apparent variability of bond strength as observed in several stretching experiments~\cite{fantner_2005,smith_1999,hansma_2005,fantner_2006,thompson_2001}. It is well known that the mechanical behavior of stretched biological molecules depends on the pulling velocity. In particular, Bell's theory~\cite{bell_1978} implies that the maximum force that a molecule can sustain varies as the logarithm of the pulling speed, as is observed in experiments~\cite{adams_2008}. Our earlier model, however, does not account for this rate dependence unless we impose the assumption that the ``random bond strength'' distribution itself varies as the logarithm of the pulling velocity. Neither does our previous model entail a recuperation of strength and toughness observed in experiments~\cite{fantner_2005}.

An understanding of these velocity- and recovery time-dependent behaviors is of paramount importance in many applications. For example, the propagation of cracks -- often caused by traumatic injuries in the case of bones -- is a dynamical process (as opposed to quasi-static). On the other hand, self-healing might impede the spread of microcracks in bone. To account analytically for these behaviors, we borrow the two-state model of protein unfolding due to Rief, Fernandez and Gaub~\cite{rief_1998}. Since breakage of sacrificial bonds and protein unfolding both involve the forced breakage of noncovalent bonds and the unraveling of compact structure, we base our framework on the assumption that the kinetics of sacrificial bond breakage can be described in a similar manner.  The two-state model enables us to obtain analytic expressions for the transition rate in terms of the force-extension profile.

The rest of this paper is organized as follows. In Sec.~II we use the two-state kinetic model to derive conditions for the breakage and formation of sacrificial bonds and show how the pulling velocity relates to these processes. In Sec.~III we apply the model to a single polymer chain -- a quasi-one-dimensional chain of entangled polymer molecules in series. We show that the model reproduces the logarithmic dependence of the peak force on the pulling velocity and, with a judicious choice of the several adjustable parameters, yields force-extension profiles that are qualitatively similar to what is observed in collagen fibril separation experiments~\cite{smith_1999,fantner_2006,hansma_2005,fantner_2005,thompson_2001,adams_2008}. In Sec.~IV we apply the kinetic model to entangled polymers and examine the effect of the delay time -- the time that permits self-healing between two successive pulling experiments -- on the macroscopic mechanical response of a pair of separating collagen fibrils. Based on these results, we propose in Sec.~V a constitutive law that describes the macroscopic response of separating collagen fibrils.

\section{Kinetic model}

In this Section we introduce a kinetic model that describes the dynamics of formation and breakage of sacrificial bonds and release of hidden length in a polymer network, and relate this to mechanical forces on the polymer network. As a simplifying assumption, we propose that the network of polymers -- in the specific example of animal bone, glue connection that hold the collagen fibrils together -- can be described as a quasi-one-dimensional ensemble of polymer chains, regardless of whether sacrificial bonds are found within a single polymer molecule or between adjacent polymers. Each polymer chain may consist of one long polymer molecule or multiple polymers entangled together; see Fig.~\ref{fig:sbhl}(a). The ensemble of these quasi-one-dimensional chains thus acquires a distribution of total lengths.

In addition, we assume that each polymer chain is semiflexible, so that the force-extension relation is given by the wormlike-chain model~\cite{bustamante_1994,marko_1995}, in conformation with much of the literature on the mechanics of proteins:
\begin{equation}\label{eq:Fx}
 F = \dfrac{k_B T}{b} \left[ \dfrac{x}{L} + \dfrac{1}{4(1 - x/L)^2} - \dfrac{1}{4} \right] .
\end{equation}
Here, the force $F$ is entropic in nature, arising from the tendency of the polymer chain to recoil and return to a state of higher entropy as it is stretched. In Eq.~\eqref{eq:Fx}, $k_B$ is the Boltzmann constant, $T$ is the temperature, $b$ is the persistence length, $x$ is the end-to-end distance, and $L$ is the contour length available for stretching -- i.e.~the total contour length of each chain minus the hidden length shielded by sacrificial bonds. Breakage of each sacrificial bond unveils hidden length, resulting in a step jump in the available contour length $L$. This results in an increase in the chain entropy which causes abrupt force drops. Stretching the chain without breaking sacrificial bonds reduces the entropy, thereby dissipating a significant amount of energy.

A sacrificial bond breaks when the force on the polymer chain exceeds the strength of that bond. As mentioned in the Introduction, we assumed in our previous work~\cite{elbanna_2012} that the bond strength is a uniform random variable, reflecting the randomness of bond breakage events. One approach to modeling the dependence of the mechanical behavior on pulling rate in the context of the previous model would be to represent the bond strength distribution itself directly as a function of pulling rate. However, this crude approach neglects the fundamental physics of rate dependence. Meanwhile, Bell's model~\cite{bell_1978} expresses the transition probability of bond formation and breakage events as a Boltzmann factor that involves the product of the force and a parameter with the dimensions of length, interpreted as the distance from the transition state of the conformational change. Along with other more sophisticated models (such as the Kramer theory~\cite{dudko_2006}), it has been successful in accounting for the rate dependence of forced protein unfolding, which in most cases involve breakage of weak internal bonds. How, then, are we to apply such kinetic models to the breakage of sacrificial bonds and release of hidden length?

We proceed by assuming that the breakage of sacrificial bonds follows a two-state pathway, so that we can apply Bell's theory. At large forces and pulling rates, the formation of sacrificial bonds can be neglected. Motivated by Su and Purohit~\cite{su_2009}, we propose that the rate of change of the number of sacrificial bonds $N_b$ is given by the first-order differential equation
\begin{equation}\label{eq:master1}
 \dfrac{d N_b^*}{dt} = - k_f N_b + k_b N_f.
\end{equation}
$N_b^*$ is the continuous version of the integer $N_b$; it represents the number of sacrificial bonds at a given instant of time, averaged over an ensemble of many polymer chains. It will be thresholded below (see Eq.~\eqref{eq:bf_internal}) to isolate individual bond breakage and formation events, and it coincides with $N_b$ whenever it is an integer. $N_f = N - 2 N_b$ is the number of free sites, with $N = L / b$ being the number of sites, or persistence lengths, in the polymer. $k_f$ and $k_b$ are the rates at which bond fragmentation and bond formation events occur; according to Bell's theory, they are given by
\begin{eqnarray}
 k_f &=& \alpha_0 \exp \left( \dfrac{F \Delta x_f}{k_B T} \right) ; \\
 k_b &=& \beta_0 \exp \left( - \dfrac{F \Delta x_b}{k_B T} \right) .
\end{eqnarray}
$F = F(x)$ is the force-extension relation given by Eq.~\eqref{eq:Fx}. $\Delta x_f$ and $\Delta x_b$ are the distances to the transition state; $\alpha_0$ and $\beta_0$ are, respectively, inverse time scales which describe the rate at which bond breakage and formation events occur at zero pulling force. We have mentioned in the Introduction that the physics of sacrificial bond breakage and protein unfolding are similar. Based on parameter estimates for the unfolding of proteins in~\cite{su_2009}, $\Delta x_f$, $\Delta x_b$ and $b$ are expected to be of the order of 0.1 nm.

A bond breakage event occurs when $N_b$ decreases by unity -- that is, when $N_b^*$ reaches an integer. Thus, the condition for a bond formation event to happen is
\begin{equation}\label{eq:bf_internal}
 \int dN_b^* = \int ( - k_f N_b + k_b N_f) dt = 1
\end{equation}
where the integral on the right hand side is over the time between successive bond formation events.
Similarly, the condition for a bond breakage event to happen is
\begin{equation}
 \int dN_b^* = \int ( - k_f N_b + k_b N_f) dt = - 1
\end{equation}
where the integral on the right hand side is over the time between successive bond breakage events. In particular, for pulling experiments at constant velocity $v$, the preceding equation gives
\begin{equation}\label{eq:bb_internal}
 \int_{x_1}^{x_2} \left( k_f \left( F(x) \right) N_b - k_b \left( F(x) \right) N_f \right) dx = v,
\end{equation}
where $x_1$ and $x_2$ are the chain end-to-end distances at successive bond breakage events.

In mechanical experiments on stretched glue connection in animal bone~\cite{fantner_2006,hansma_2005,fantner_2005} it has been found that sacrificial bonds mediated by ions such as calcium also form between the glue strand backbone and the collagen fibrils. Breakage of these end bonds causes the detachment of the entire glue strand from one of the collagen fibrils, so that the stretching force on the glue strand immediately drops to zero. In addition, it has been found that broken links may self-heal, in that some broken end bonds would be restored if, after a particular pulling experiment, the entire sample is left untouched for times as short as a few seconds~\cite{fantner_2005}. We propose that the breakage and restoration of end bonds can be described within the same theoretical framework. That is, the change in the number of end bonds $N_e$ is governed by the rate equation
\begin{equation}\label{eq:endbond}
 \dfrac{dN_e^*}{dt} = - k_f^{end} N_e + k_b^{end} (1 - N_e).
\end{equation}
$N_e^*$, the continuous version of the integer $N_e$, varies between zero and unity; it can be interpreted as the fraction of end bonds that have yet to break. Notice that $(1 - N_e)$ appears because each glue strand either attaches to the bone fibril, or does not anchor to it. $k_f^{end}$ and $k_b^{end}$ are the rates of end bond breakage and formation, necessarily different from $k_f$ and $k_b$, given by
\begin{eqnarray}
 k_f^{end} &=& \alpha_e \exp \left( \dfrac{F \Delta x_f^{end}}{k_B T} \right), \\
 k_b^{end} &=& \beta_e \exp \left( - \dfrac{F \Delta x_b^{end}}{k_B T} \right).
\end{eqnarray}
As before, $\alpha_e$ and $\beta_e$ are respectively the rates at which end bonds break and form when no external force is present, and $\Delta x_f^{end}$ and $\Delta x_b^{end}$ are the distances from the transition state for end bond breakage and formation events. For a single polymer chain, the end bond breaks when
\begin{equation}\label{eq:bb_end}
 \int_0^{x_c} k_f^{end} (F(x)) dx = v
\end{equation}
where $x_c$ is the chain end-to-end distance at which the end bond breaks and the chain detaches.

We note on passing that for a collection of polymers stacked in parallel, Eq.~\eqref{eq:endbond} should more properly be interpreted as the governing equation for the fraction of polymer chains with end bonds restored as a function of time $t$; thus,
\begin{equation}\label{eq:master2}
\dfrac{dN_e^*}{dt} = - k_f^{end} N_e^* + k_b^{end} (1 - N_e^*).
\end{equation}
Suppose $t = 0$ marks the time at which all polymers detach from the surface, after the previous stretching experiment. Then the fraction $N_e^*$ of polymers that adhere to the surface at time $t$ is given by
\begin{equation}\label{eq:restore}
 N_e^* = \dfrac{\beta_e}{\alpha_e + \beta_e} \left[ 1 - e^{- (\alpha_e + \beta_e) t} \right] .
\end{equation}

Equations~\eqref{eq:bb_internal} and~\eqref{eq:bb_end} can be used to predict the force-extension curve of a stretched polymer; in particular, they can predict the chain end-to-end distance at which bond breakage events occur, as well as the corresponding bond strengths. Meanwhile, Eq.~\eqref{eq:restore} is particularly useful for analyzing the dependence of the mechanical behavior of multiple polymers stacked in parallel on the delay time between pulls.

Note that while our model is deterministic and does not capture the randomness of bond breakage events as in~\cite{elbanna_2012}, it represents an average over a large ensemble of experiments.

\section{Pulling a single polymer chain: theoretical predictions}

We begin by considering the force-extension behavior of a single polymer chain whose total length is $L_c = 100$ nm. Let $m$ denote the number of sacrificial bonds. For simplicity, we assume that shielded lengths do not contain sacrificial bonds, and that the length $L_j$ of each hidden loop is a uniform random number less than $L_c / m$. Then the initial available length is $L_i = L_c - \sum_{j=1}^m L_j$. To locate bond breakage events, we integrate Eq.~\eqref{eq:bb_internal} over the force-extension profile, and compute the end-to-end distance $x_2$ at which each individual bond breakage event occurs, assuming that bonds break in the order of increasing shielded length. We integrate Eq.~\eqref{eq:bb_end} over the entire force-extension curve to locate the maximum pulling distance before the polymer chain detaches from the underlying material.

\begin{figure}[here]
\centering 
\subfigure{\includegraphics[width=.45\textwidth]{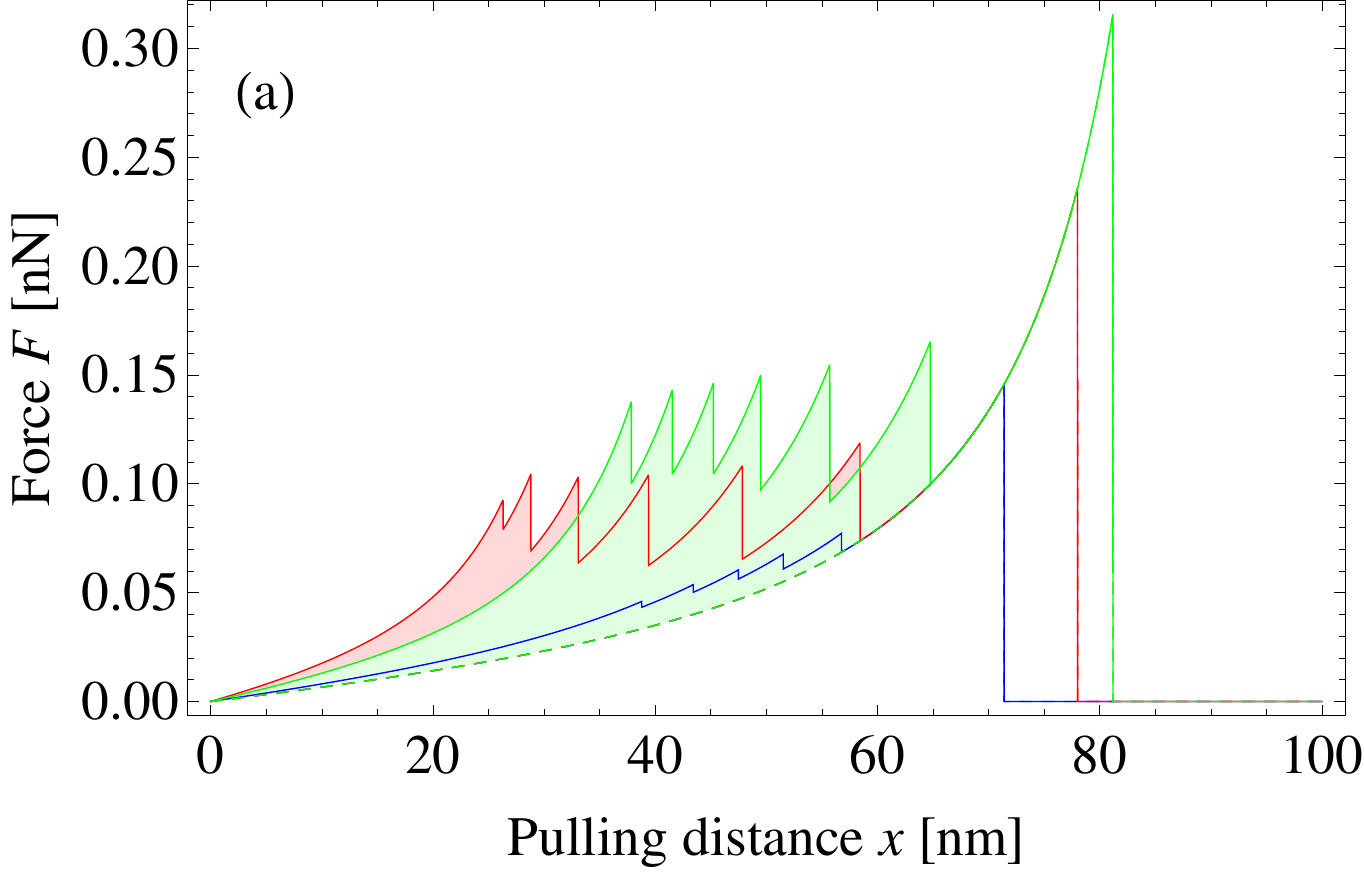} \label{fig:fxplot}}
\subfigure{\includegraphics[width=.45\textwidth]{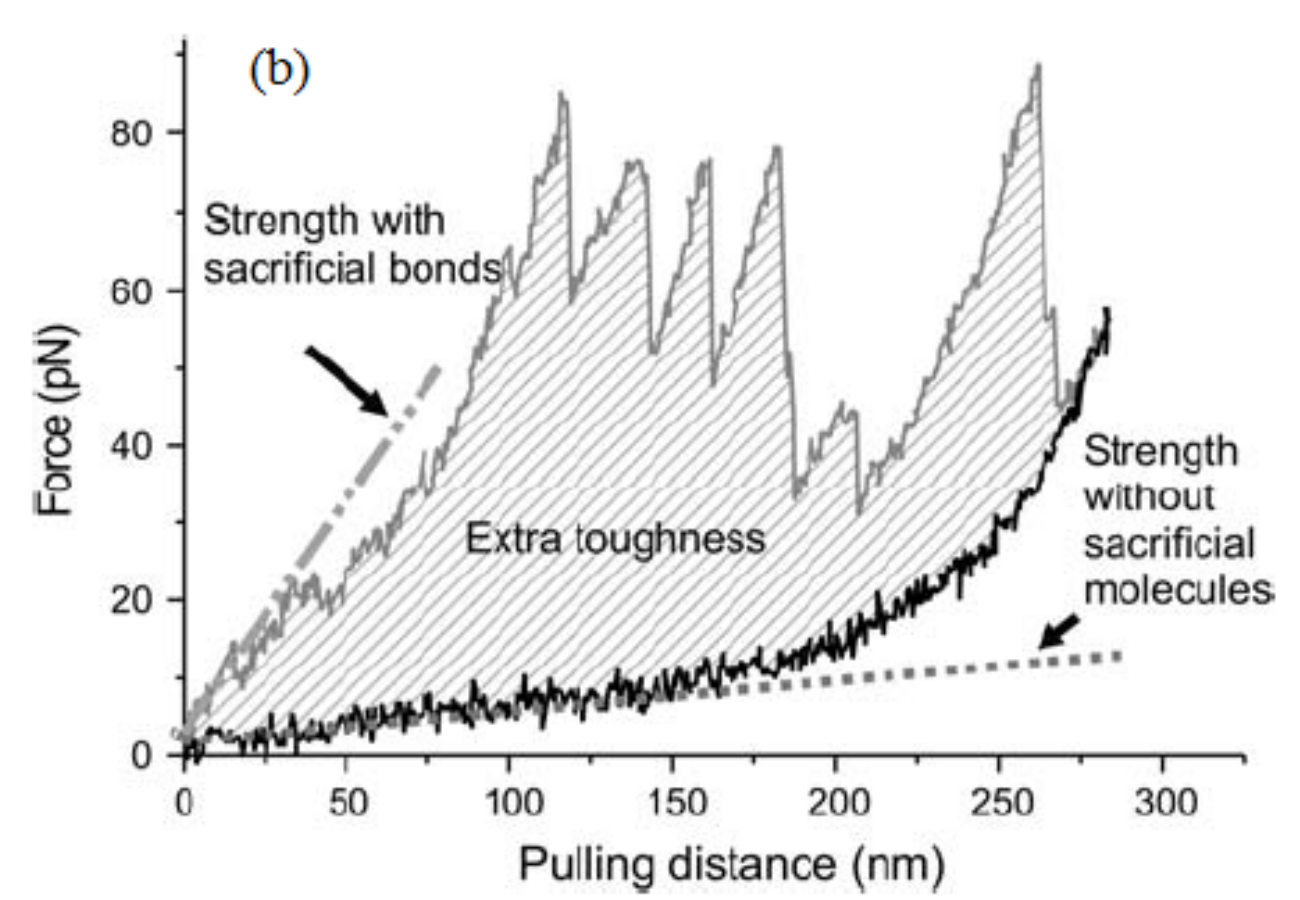} \label{fig:inset_fxplot}}
\caption{(Color online) (a) Force-extension curves of a polymer with a total contour length of $L_c = 100$ nm and with $m = 6$ sacrificial bonds, stretched at $v = 10^2$ nm s$^{-1}$ (blue), $10^3$ nm s$^{-1}$ (red), and $10^4$ nm s$^{-1}$ (green), increasing from bottom to top. (b) Comparison between sample experimental pulling curves of a polymer chain with sacrificial bonds and hidden length (upper grey curve) and one without (lower black curve). Adapted from~\cite{fantner_2006}. In both figures, the shaded area represents the extra energy dissipated through sacrificial bonds, a measure of the increase in toughness of the material.} 
\end{figure}

Figure~\ref{fig:fxplot} show the force-extension curves of a polymer chain with $m = 6$ sacrificial bonds, stretched at four representative velocities $v = 10^2, 10^3$ and $10^4$ nm s$^{-1}$. In computing these theoretical curves we have used the parameter estimates $\alpha_0 = 0.3$ s$^{-1}$, $\beta_0 = 0.003$ s$^{-1}$, $\alpha_e = 0.1$ s$^{-1}$, $b = 0.1$ nm, $\Delta x_f = 0.25$ nm, $\Delta x_b = 0.1$ nm, and $\Delta x_e = 0.15$ nm. The magnitudes of these parameters are roughly consistent with those in protein unfolding models (for example~\cite{su_2009}). The average internal bond strength varies from roughly 80 pN at $v = 10^2$ nm s$^{-1}$ to 150 pN at $v = 10^4$ nm s$^{-1}$, while the end bond strength varies from roughly 150 pN to 300 pN over the same range of pulling velocities. For comparison, Fig.~\ref{fig:inset_fxplot} shows sample experimental pulling curves of a polymer chain with sacrificial bonds and hidden length. The force drops due to breakage of sacrificial bonds and release of hidden length are also seen in the grey curve here, which represents the behavior of a typical polymer chain with sacrificial bonds, stretched by the tip of an atomic force microscope. The black curve shows the mechanical response of an otherwise identical polymer chain, i.e., one with the same total length but with no sacrificial bonds. In both figures, the shaded area between the two curves represents the increase in toughness due to the presence of sacrificial bonds and hidden length. Due to the inherent variability of experimental parameters (ion concentration in buffer, chain length, etc.)~in different samples, the aim of this comparison with individual chain pulling experiments is to seek qualitative, rather than exact quantitative, agreement; qualitative agreement between the theoretical and experimental results is clear.

\begin{figure}
\centering
\subfigure{\includegraphics[width=.45\textwidth]{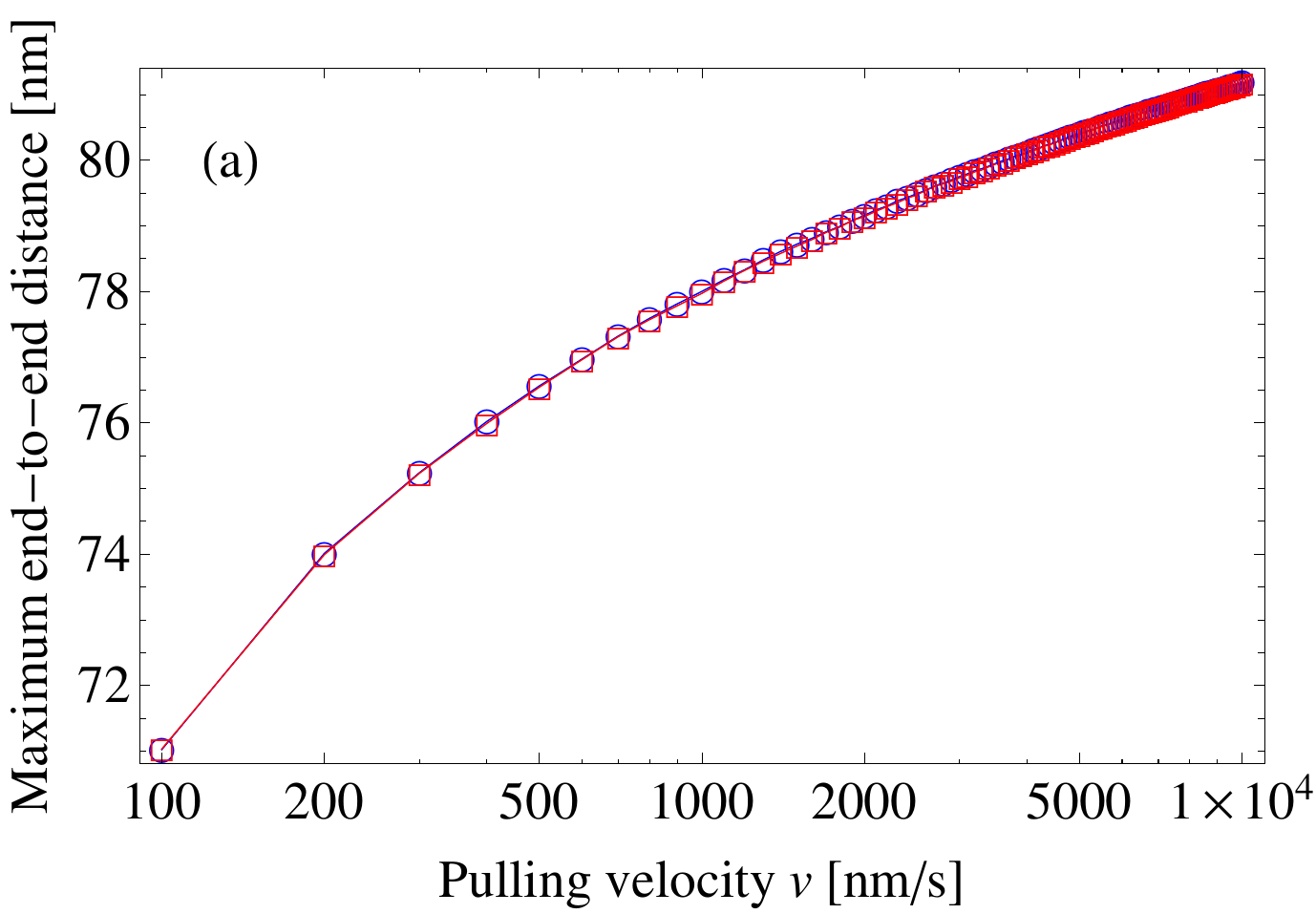} \label{fig:xvplot}}
\subfigure{\includegraphics[width=.45\textwidth]{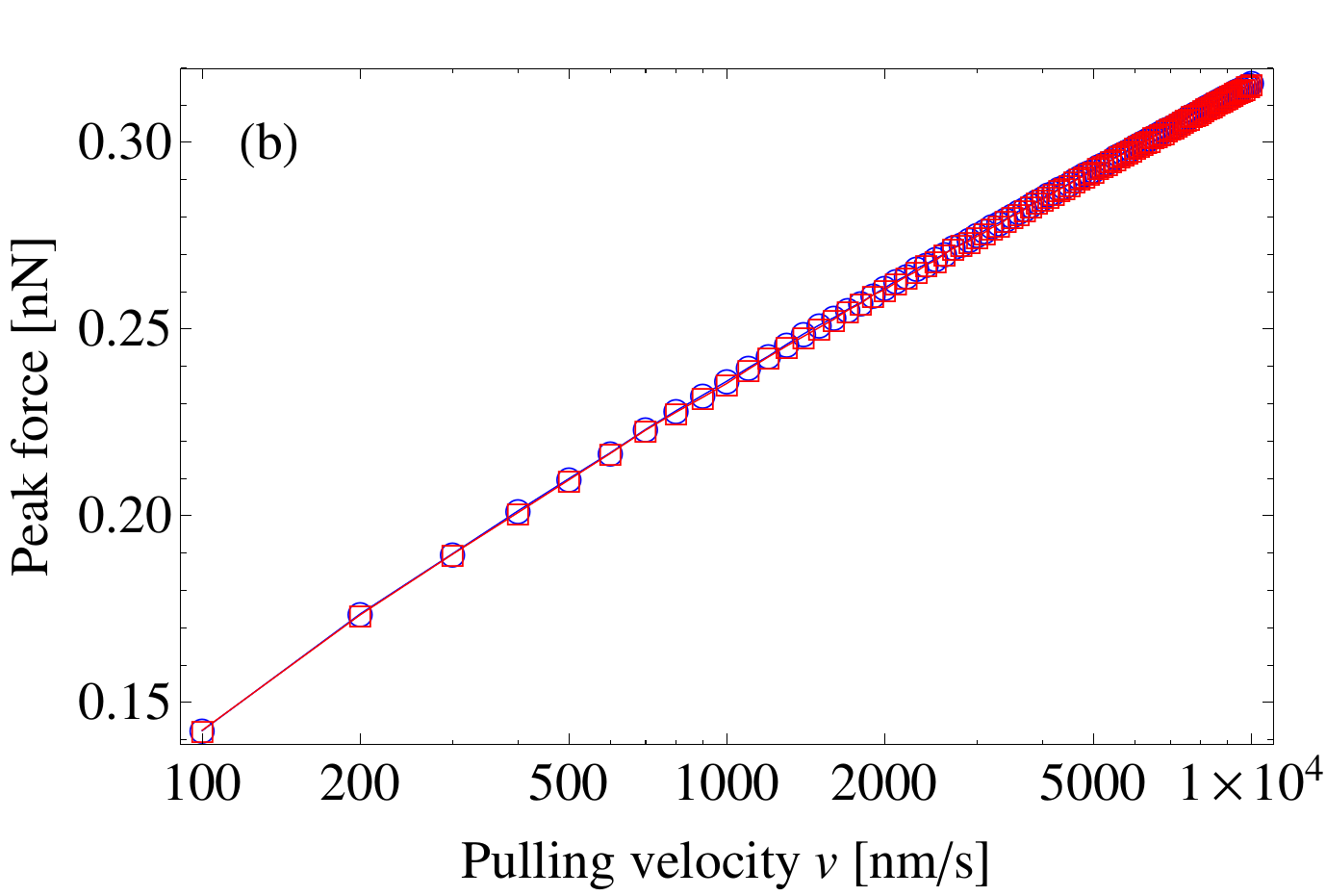} \label{fig:fvplot}}
\subfigure{\includegraphics[width=.45\textwidth]{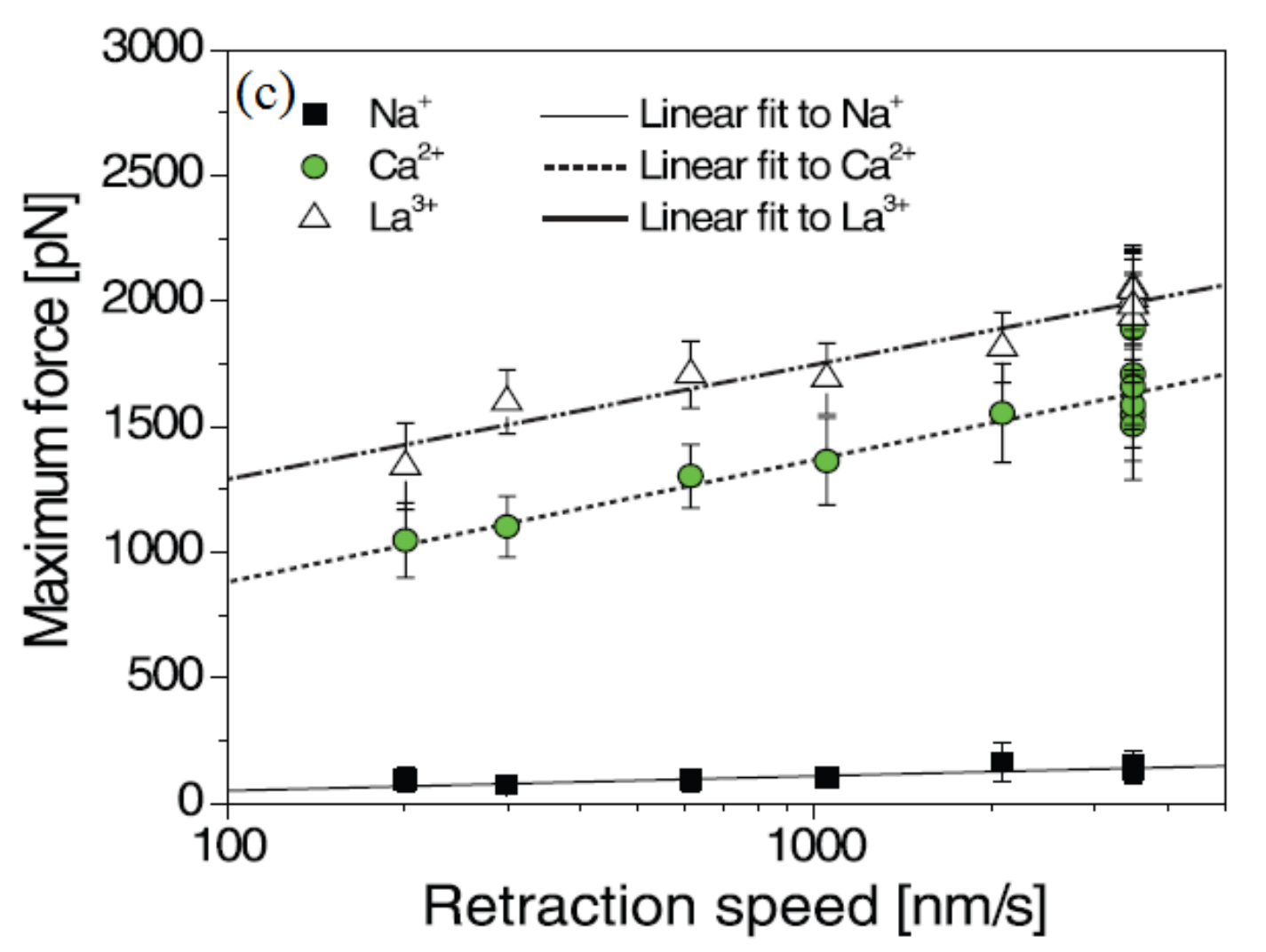} \label{fig:inset_fvplot}}
\caption{(Color online) (a) Maximum pulling distance, and (b) end bond strength, as functions of pulling velocity $v$, for $m = 6$ (blue circles) and 12 (red squares) sacrificial bonds. The overlap of the curves indicates that both quantities appear to be independent of $m$. Figure (c) is the dynamic force spectrum of dentin matrix protein 1 with sacrificial bonds mediated by sodium, calcium and lanthanum buffers respectively, adapted with permission from~\cite{adams_2008}. Our theory predicts a log-linear dependence, as observed in experiments, of the end bond strength on the pulling rate.}
\end{figure}

Figure~\ref{fig:xvplot} shows the variation of the maximum stretch $x_c$ as a function of the pulling velocity, for two polymer chains with 6 and 12 sacrificial bonds respectively but are otherwise identical; the overlap of the curves indicates that this is independent of the number of sacrificial bonds. Figure~\ref{fig:fvplot} is a log-linear plot that displays the variation of the end bond strength as a function of the pulling velocity $v$. The linearity of the plot shows that the end bond strength, which equals the maximum stretching force along the stretching profile, varies logarithmically with the stretching velocity $v$, as is predicted by Bell's theory~\cite{bell_1978} and seen in experiments with dentin matrix protein 1~\cite{adams_2008}, shown here in Fig.~\ref{fig:inset_fvplot}. 

\begin{figure}[here]
\centering \includegraphics[width=.45\textwidth]{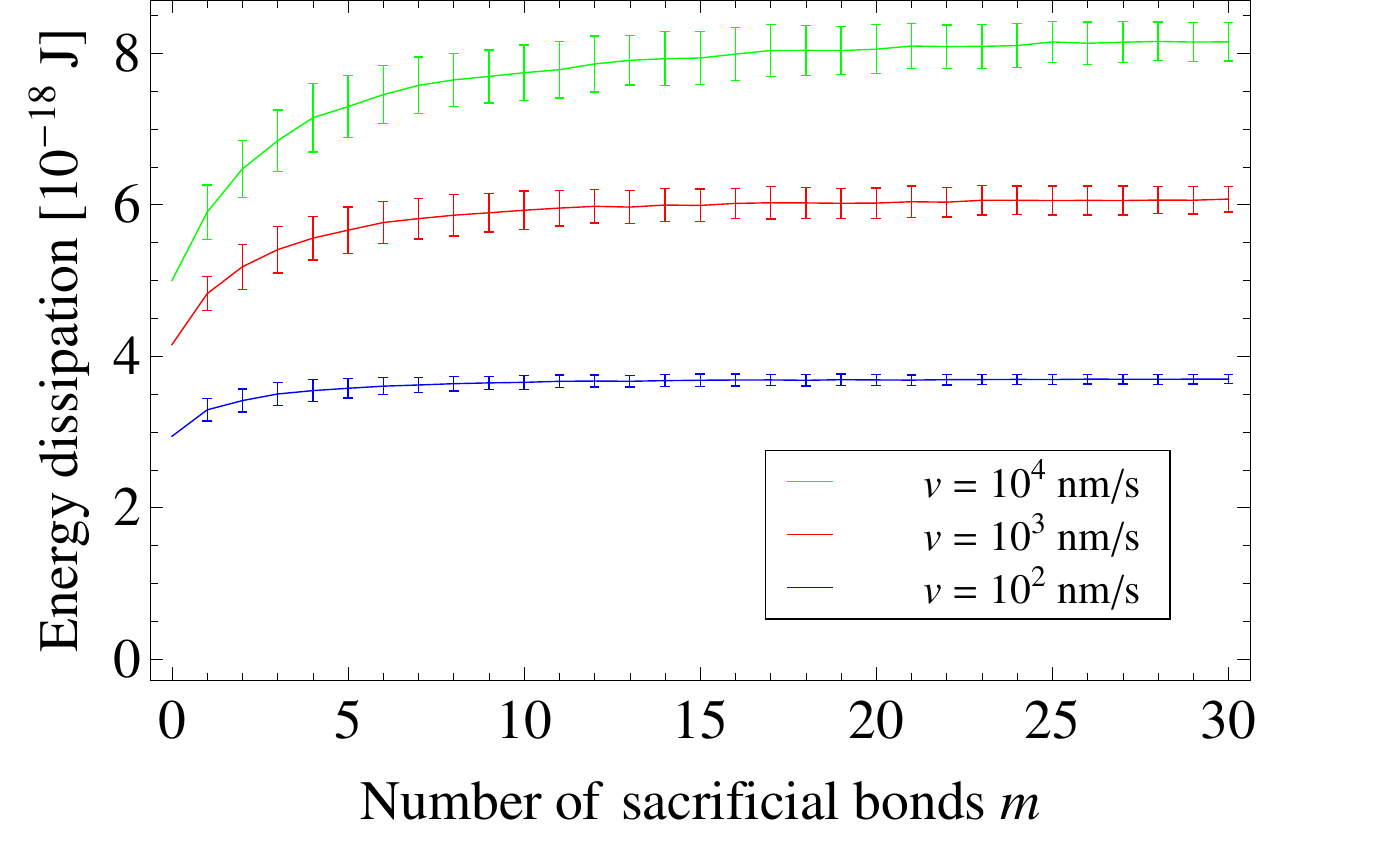} \caption{\label{fig:emplot}(Color online) Total energy dissipation over the entire course of stretching, as a function of the number of sacrificial bonds $m$, for stretching velocities $v = 10^2$ nm s$^{-1}$ (blue), $10^3$ nm s$^{-1}$ (red), and $10^4$ nm s$^{-1}$ (green), increasing from bottom to top. Results are averaged over 200 runs. The vertical bars indicate one standard deviation.} 
\end{figure}

Figure~\ref{fig:emplot} shows the total energy dissipation, a measure of the toughness of the polymer chain given by the area under the force-extension curve, as a function of the number of sacrificial bonds $m$, for three representative stretching velocities $v = 10^2, 10^3$ and $10^4$ nm s$^{-1}$, and averaged over 200 runs. While sacrificial bonds and hidden length constitute a major toughening mechanism, increasing the number of sacrificial bonds beyond $m \approx 15$ fails to further stiffen the chain, as is found in~\cite{elbanna_2012}. In addition, the importance of this toughening mechanism becomes more pronounced at high pulling velocities (of the order of 1000 nm s$^{-1}$), providing increased resistance against impact loading.

\section{Stretching multiple polymer chains in parallel; effect of delay time between pulls}
 
In this section we consider the dynamical behavior of multiple polymers stretched in parallel. For simplicity, assume that each polymer chain is independent of the others, with no entanglement between the polymer strands, so that the total force equals the sum of forces in each polymer chain.

\begin{figure}[here]
\centering \includegraphics[width=.45\textwidth]{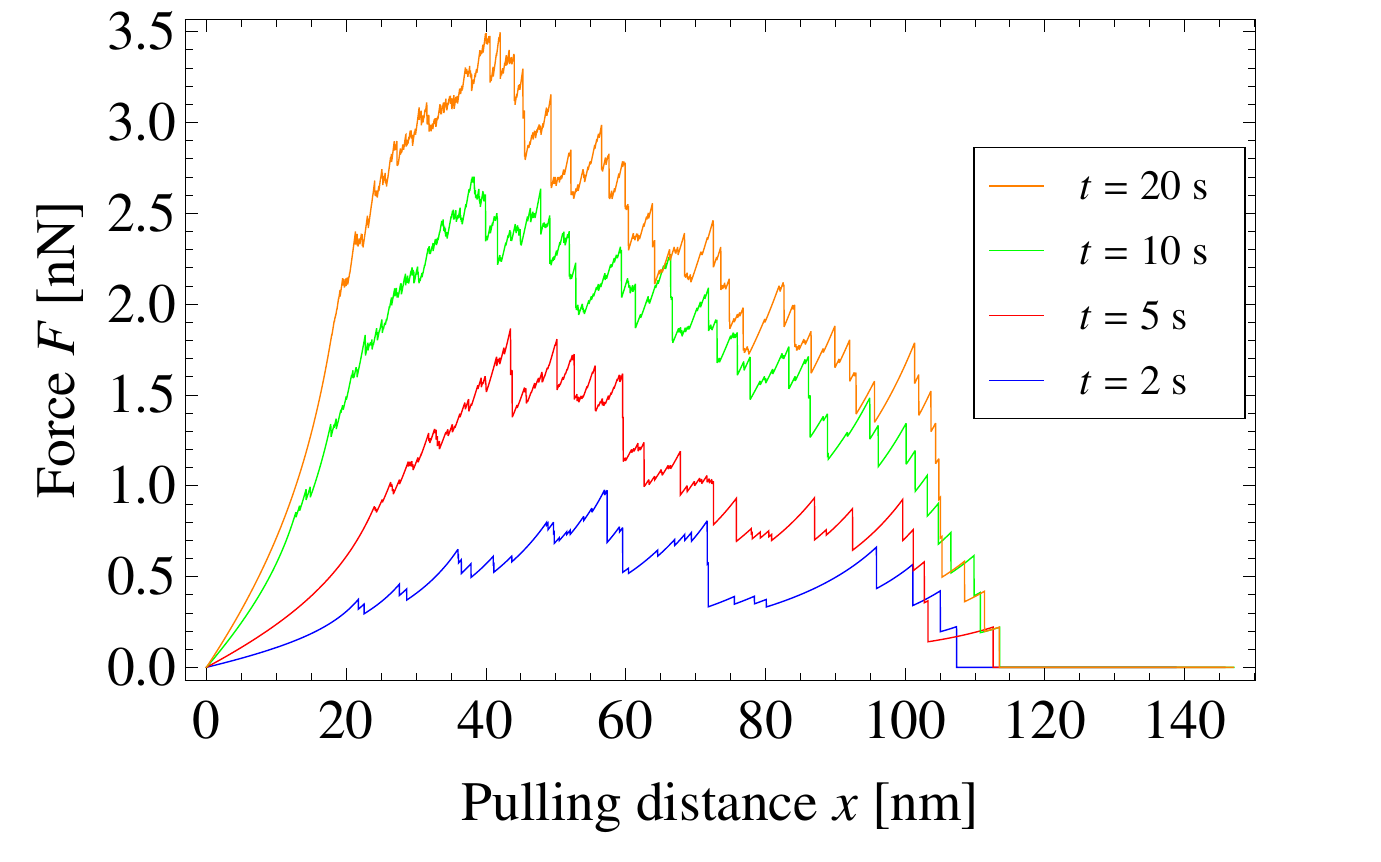} \caption{\label{fig:stackedfxplot}(Color online) Force-stretch curves for $N_p = 200$ polymer chains stacked in parallel. The delay times are $t = 2$ s (blue), 5 s (red), 10 s (green) and 20 s (orange) from bottom to top, corresponding $N_p N_e^* = 8$, 18, 28 and 36 polymer chains adhering to both pieces of substrate (collagen fibril) at the beginning (see Eq.~\eqref{eq:restore} for an expression for $N_e^*$).} 
\end{figure}

Figure~\ref{fig:stackedfxplot} shows the force-extension curves of $N_p = 200$ parallel polymer chains, at pulling velocity $v = 1000$ nm s$^{-1}$, for delay times (the time between rupture of all end bonds and the start of the next pulling experiment) ranging from 1 to 20 seconds. In computing these theoretical curves we have used Eq.~\eqref{eq:restore} to calculate the fraction of polymers with restored end bond connections to the underlying substrate (for example, mineralized collagen fibrils in the case of glue connection) as a function of the delay time $t$. We assume that the total contour length $L_c$ of each polymer is uniformly and randomly distributed between 50 and 150 nm. We use Eq.~\eqref{eq:bf_internal} to compute the number of internal sacrificial bonds that form as a function of the delay time, and assign random hidden lengths to each of these sacrificial bonds as before. As before, we have chosen $\alpha_0 = 0.3$ s$^{-1}$, $\beta_0 = 0.003$ s$^{-1}$, and $\alpha_e = 0.1$ s$^{-1}$. To account for the relatively slow recovery of ruptured polymer chains as a function of the delay time, as seen in~\cite{fantner_2005}, we choose $\beta_e = 0.025$ s$^{-1}$. These sample pulling curves indicate that the extension at maximum force and the maximum extension at a given pulling velocity is independent of the delay time. Also, the force peaks level off for large delay times. This follows from Eq.~\eqref{eq:restore}, which shows that the fraction of polymers with restored end bond connections approaches the asymptotic limit $\beta_e / (\alpha_e + \beta_e)$ as the delay time $t$ becomes large; only those polymers with restored end bonds carry the pulling force and contribute to energy dissipation. In the limit $\alpha_e \ll \beta_e$, all end bonds are restored for large delay times between successive pulling experiments. Our present choice of parameters, however, stipulates that at most one-eighth of all glue strands are attached to bone fibrils at both ends.

\begin{figure}[here]
\centering 
\subfigure{\includegraphics[width=.45\textwidth]{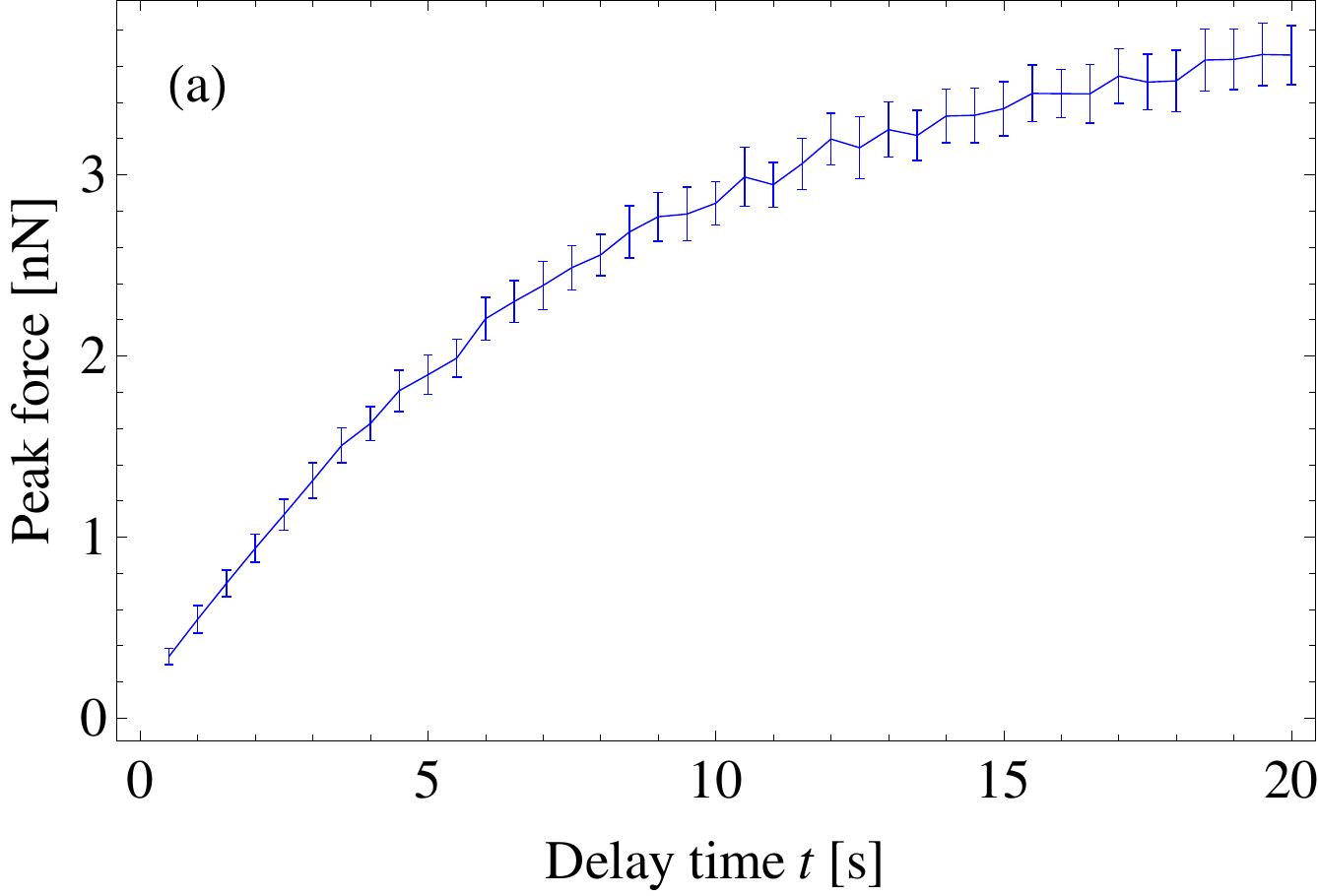} \label{fig:stackedftplot}}
\subfigure{\includegraphics[width=.45\textwidth]{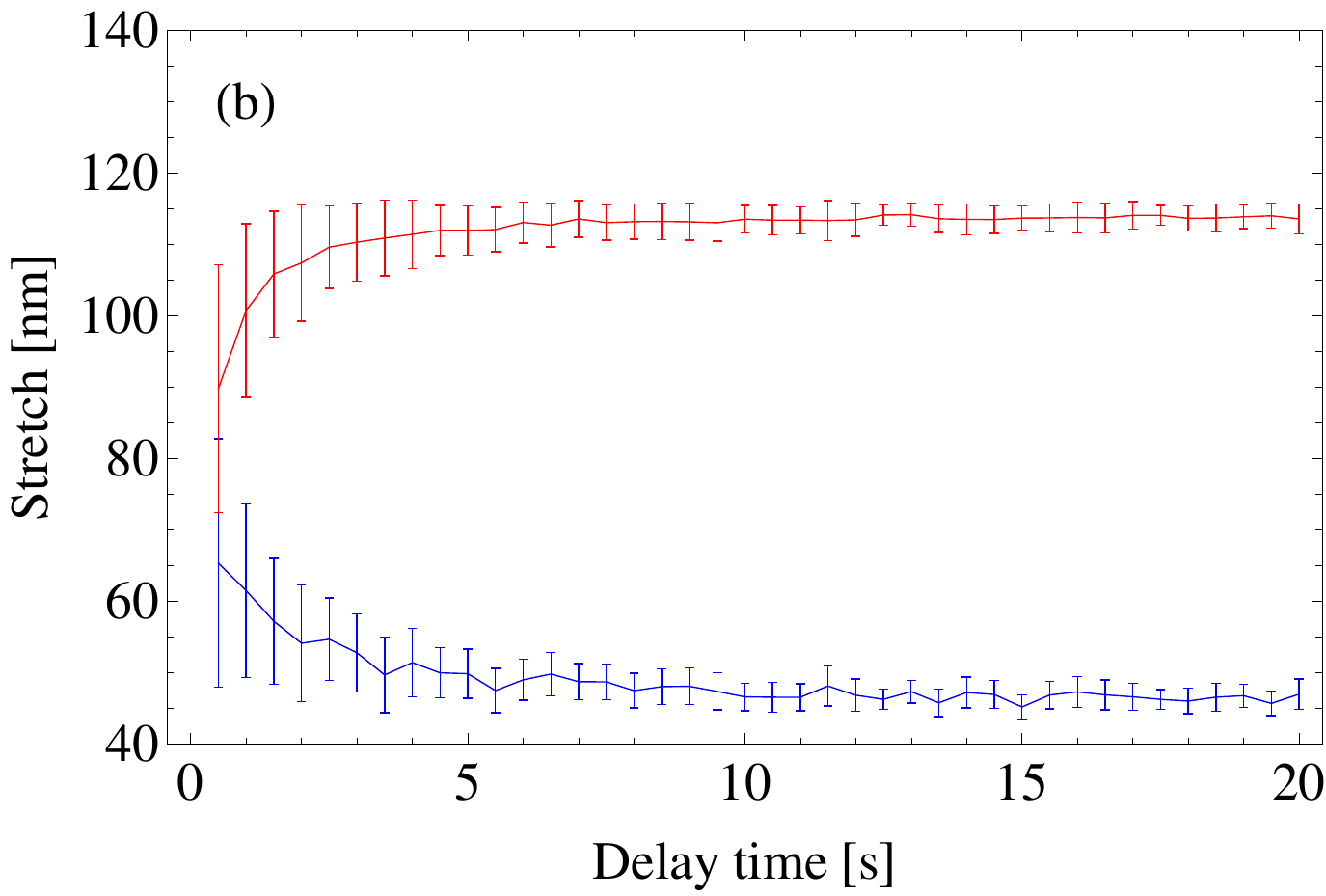} \label{fig:stackedxtplot}}
\caption{(Color online) (a) Peak force, and (b) displacement at maximum force (blue, bottom curve) and maximum extension (red, top curve), as functions of delay time, for $N_p = 200$ polymers stacked in parallel, pulled at $v = 1000$ nm s$^{-1}$. Results are averaged over 100 runs, and the error bars indicate one standard deviation.} 
\end{figure}

To verify these claims, Fig.~\ref{fig:stackedftplot} shows the peak force as a function of delay time $t$ between pulls, for $N_p = 200$ parallel polymers pulled at a velocity $v = 1000$ nm s$^{-1}$, averaged over 100 runs. The peak force increases roughly in proportion to $N_e^* = (\beta_e / (\alpha_e + \beta_e) ) (1 - e^{- (\alpha_e + \beta_e) t} )$, the fraction of polymers that possess restored end connections and therefore transmit the force. Figure~\ref{fig:stackedxtplot} shows the displacement at maximum stretching force, as well as the maximum stretch, as a function of delay time between pulls. Both quantities are roughly independent of the delay time, except that the displacement at maximum stretching force shows a slight decrease for small delay times. This can be traced to the fact that the small number of internal sacrificial bonds that restore for small delays times leads to fewer cusps in the force-extension curve of each polymer chain and reduces the extra energy dissipation brought by these sacrificial structures (see below), thereby delaying the occurrence of the maximum force.

\begin{figure}[here]
\centering 
\subfigure{\includegraphics[width=.45\textwidth]{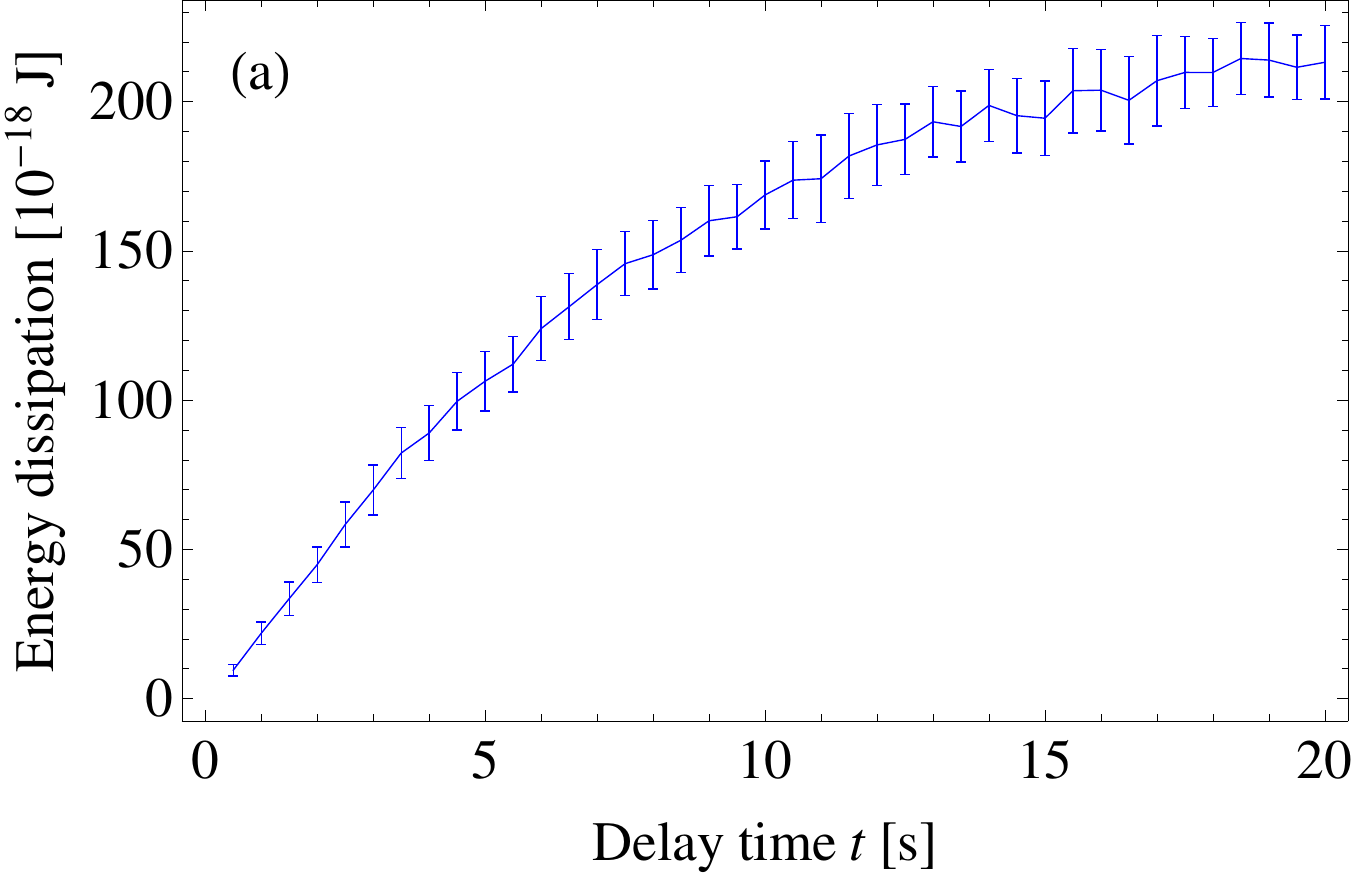} \label{fig:stackedetplot}}
\subfigure{\includegraphics[width=.45\textwidth]{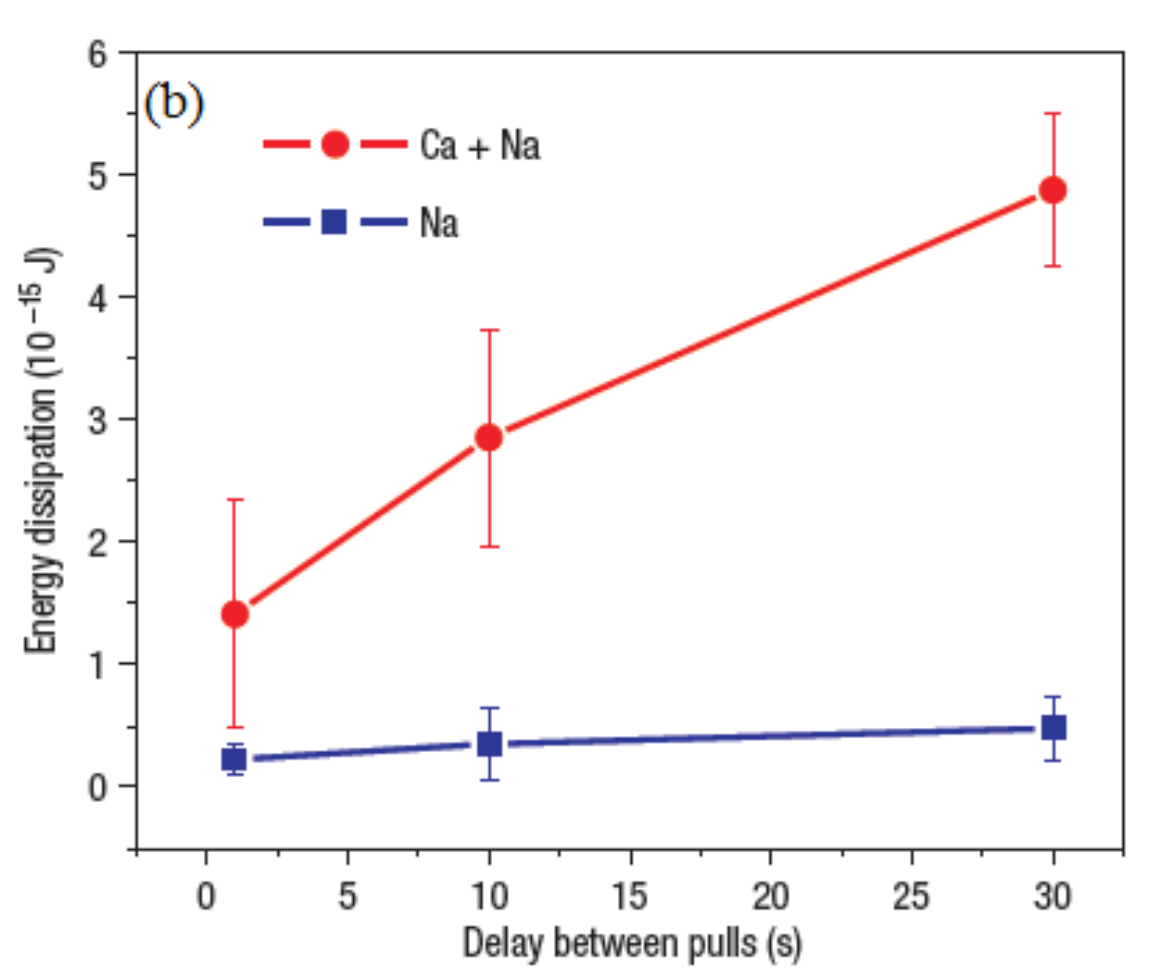} \label{fig:inset_stackedetplot}}
\caption{(Color online) (a) Total energy dissipation over the entire course of stretching, as a function of delay time, for $N_p = 200$ polymers stacked in parallel, pulled at $v = 1000$ nm s$^{-1}$. Results are averaged over 100 runs, and the error bars indicate one standard deviation. (b) Figure 2(c) from~\cite{fantner_2005}, reproduced here for comparison, shows the energy dissipation involved in the separation of bone fibrils, in a buffer where calcium and sodium ions are present (red), and in a buffer where only sodium ions are present (blue). The authors there concluded that calcium ions lead to enhanced bond strength, and within our choice of parameters, our theoretical prediction for the energy dissipation qualitatively matches that of~\cite{fantner_2005} in the presence of calcium ions.} 
\end{figure}

Figure~\ref{fig:stackedetplot} shows the total energy dissipation, a measure of the toughness of the glue connection between the two pieces of underlying material, as a function of the delay time $t$ between pulls. For small delay times, the marked growth in the number of restored end bond connections, the number of restored internal sacrificial bonds, and the increase in energy dissipation as a function of the number of internal sacrificial bonds (cf.~Fig.~\ref{fig:emplot}) all contribute to the fast increase of total energy dissipation as a function of the delay time. For large delay times, however, these growths slow down gradually, so that the increase in energy dissipation flattens out. A comparison to Fig.~\ref{fig:inset_stackedetplot}, which shows the experimental measurements for the energy dissipation involved in the separation of two bone fibrils~\cite{fantner_2005}, indicates that our theoretical prediction qualitatively matches the experimental observations of bone fibrils in a buffer of calcium and sodium ions. The result for a sodium buffer corresponds to different choices for the rate parameters $k_f$, $k_b$, $\alpha_0$ and $\beta_0$, etc.

\section{Towards a constitutive law for force as a function of distance}

In multiscale simulations of bone fracture it is necessary to incorporate a constitutive law for the separation of collagen fibrils under tensile stress. More specifically, in realistic situations where hundreds of glue strands are present between each pair of collagen fibrils, we need, in a mean-field sense (i.e.~smoothing out all abrupt force drops due to bond breakage or detachment), a force law $F(x, v, t)$ for the force on the polymeric system under stretch.

\begin{figure}[here]
\centering \includegraphics[width=.45\textwidth]{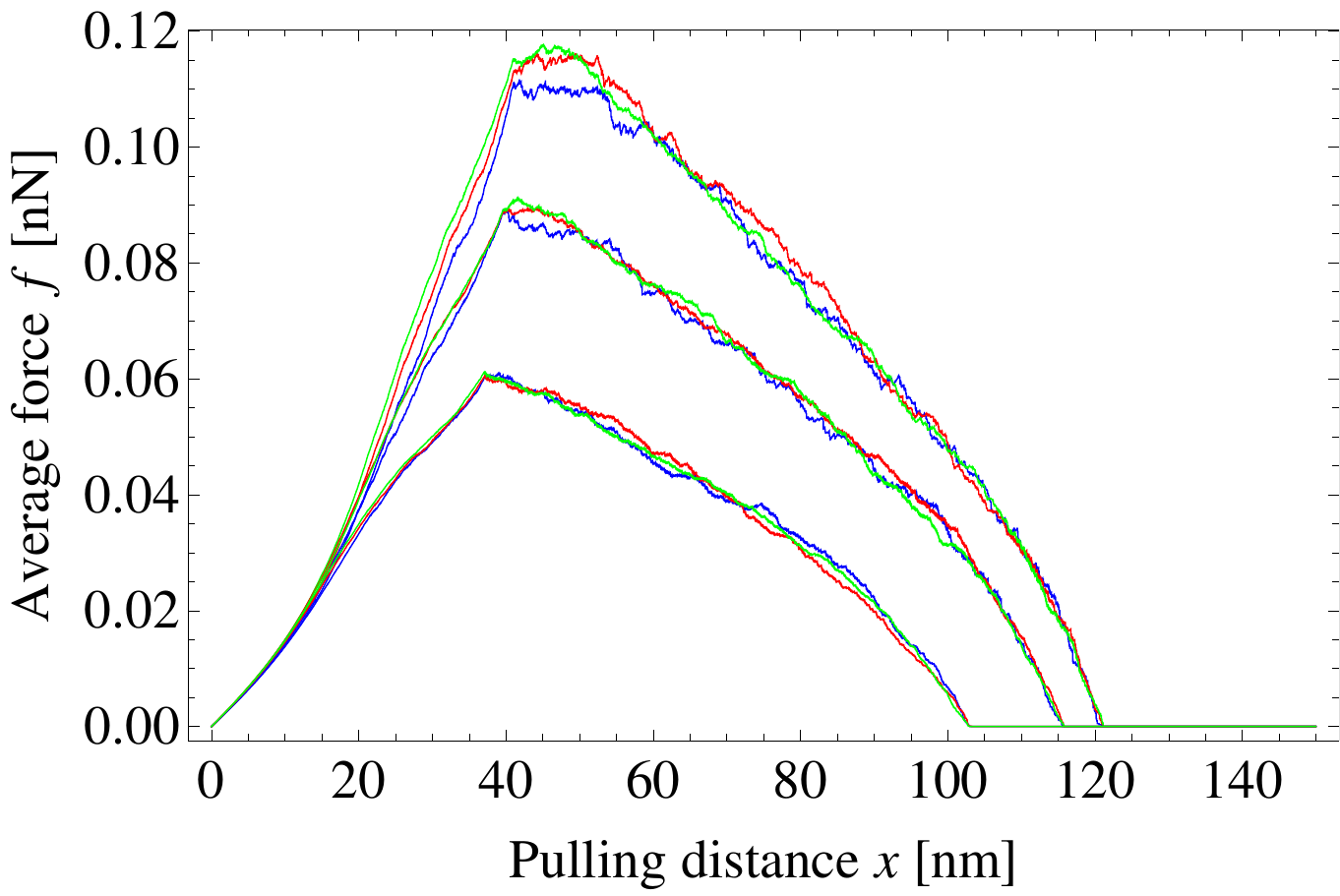} \caption{\label{fig:fxsmoothplot}(Color online) Force per polymer chain that is attached at both ends at the beginning of the experiments, as a function of pulling distance, for $N_p = 200$ polymers. Results are averaged over 100 runs, thereby smoothing out abrupt force drops. The delay times are $t = 5$ s (blue), 10 s (red) and 20 s (green). The pulling rates are $v = 100$, 1000 and 10000 nm s$^{-1}$ for the three sets of curves, from bottom to top.} 
\end{figure}

\begin{figure}[here]
\centering \includegraphics[width=.45\textwidth]{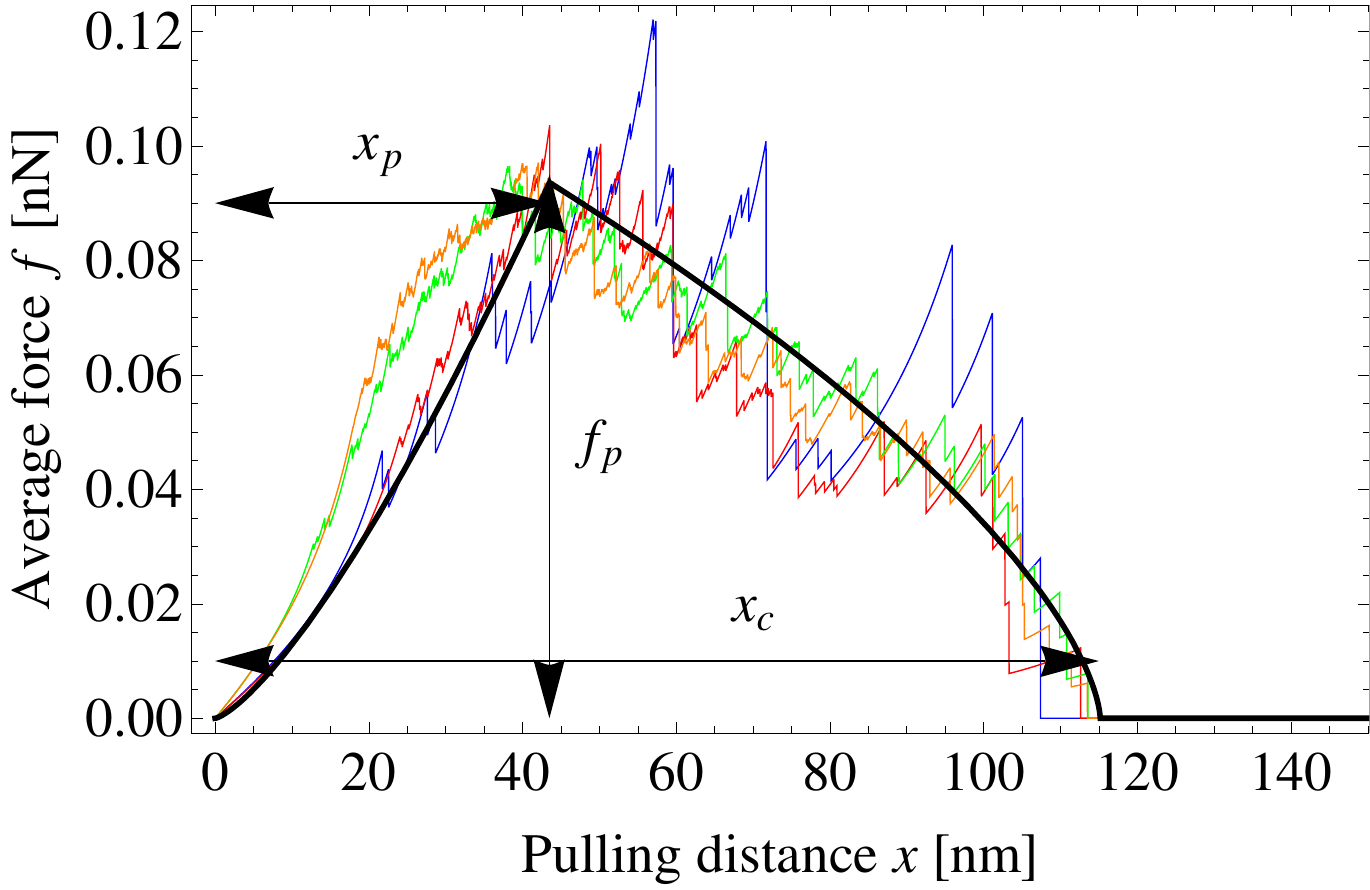} \caption{\label{fig:fxfit}(Color online) Force-stretch curves as in Fig.~\ref{fig:stackedfxplot} for $N_p = 200$ polymers stacked in parallel, pulled at $v = 1000$ nm s$^{-1}$, normalized by the number of polymer chains that are attached at the beginning of the experiment. The delay times are $t = 2$ s (blue), 5 s (red), 10 s (green) and 20 s (orange). The thick black curve represents the fitting function Eq.~\eqref{eq:fxv} with parameter values given in the text. Note that the blue curve for the short delay time $t = 2$ s is noticeably jagged, owing to the fact that only 8 polymer chains (see Eq.~\eqref{eq:restore}) adhere to the bone fibrils at the beginning. The quantities $f_p$, $x_p$ and $x_c$ appearing in Eq.~\eqref{eq:fxv} are labeled in the figure.} 
\end{figure}

To this end, Fig.~\ref{fig:fxsmoothplot} shows sample force-extension curves of $N_p = 200$ polymer chains stacked in parallel, normalized by the number of chains that are attached at both ends at the beginning, ignoring all interactions between them, for delay times ranging between 1 and 10 seconds. In computing these theoretical curves we have averaged over 100 pulling experiments and divided the total force $F(x, v, t)$ by the total number of polymer chains $N_p N_e^*$ that are attached to the collagen fibrils at both ends at the beginning, where $N_e^* = (\beta_e / (\alpha_e + \beta_e))(1 - e^{-(\alpha_e + \beta_e) t})$ as given by Eq.~\eqref{eq:restore}. As in Fig.~\ref{fig:stackedfxplot} the total contour length $L_c$ of each polymer is uniformly distributed between 50 and 150 nm. Importantly, for each pulling velocity $v$, the curves for different delay times $t$ collapse together. This implies that in the limit of long delay times $t$, the total force in separating two pieces of collagen fibril is given as a function of distance $x$ and separation velocity $v$ by
\begin{equation}
 F(x, v, t) = N_p N_e^*(t) f(x, v) .
\end{equation}
The delay time dependence comes in only through the fraction $N_e^*(t)$ of polymer chains attached to the bone fibrils at the beginning. The average force on each of these polymer chains, $f(x, v)$, is independent of the delay time $t$ and can be approximated by separate power law fits to the increasing (strengthening) and decreasing (weakening) portions of the curves:
\begin{equation}\label{eq:fxv}
 f(x, v) =
 \begin{cases}
 f_p (v) \left( \dfrac{x}{x_p(v)} \right)^{s_1}, & \text{if $x \leq x_p(v)$} ; \\
 f_p (v) \left( \dfrac{x_c(v) - x}{x_c(v) - x_p(v)} \right)^{s_2}, & \text{if $x_p(v) < x < x_c(v)$} ; \\
 0, & \text{if $x \geq x_c(v)$}.
 \end{cases}
\end{equation}
Here, $f_p (v)$, $x_p (v)$ and $x_c(v)$ are the velocity-dependent peak force, end-to-end distance at peak force, and maximum pulling distance, respectively. We find that $s_1 \approx 1.35$ and $s_2 \approx 0.65$ and, in the case $v = 1000$ nm s$^{-1}$ shown in Fig.~\ref{fig:fxsmoothplot}, that $f_p \approx 0.094$ nN, $x_p \approx 43.5$ nm and $x_c \approx 115$ nm. Figure~\ref{fig:fxfit} shows how this function fits sample force-displacement profiles.

\begin{figure}[here]
\centering \includegraphics[width=.45\textwidth]{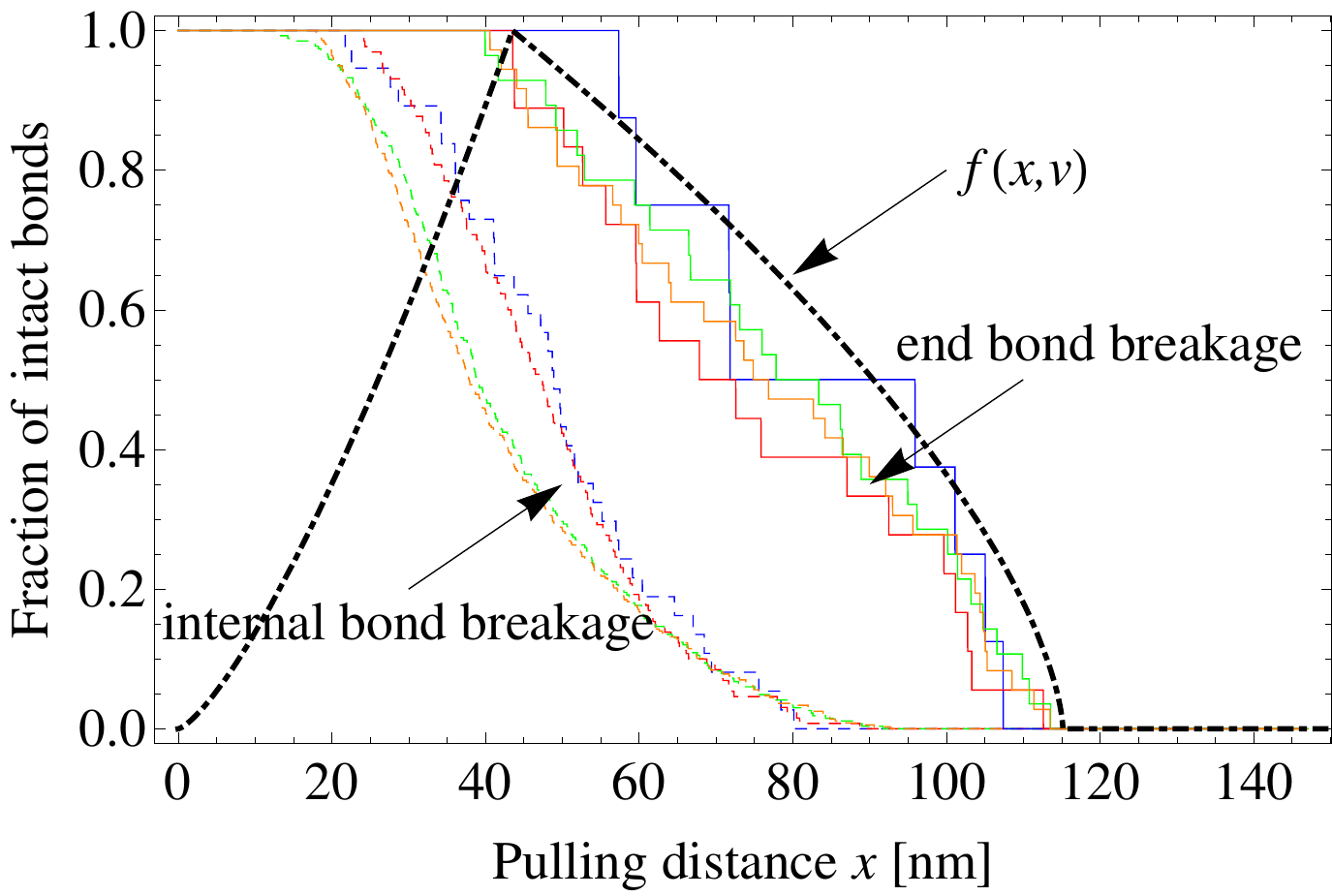} \caption{\label{fig:nxplot}(Color online) Number of polymer chains that have yet to rupture, for $N_p = 200$ polymers stacked in parallel, pulled at $v = 1000$ nm s$^{-1}$, normalized by the number of polymer chains attached to bone fibrils at the beginning (solid curves); and fraction of internal sacrificial bonds that have yet to rupture (dashed curves). The delay times are $t = 2$ s (blue), 5 s (red), 10 s (green) and 20 s (orange). The black dot-dashed curve is a vertically rescaled $f(x, v)$ and shows that the onset of polymer chain detachment coincides with the transition to weakening behavior after reaching the maximum force $f_p$. This illustrates the interplay between microscopic physics (bond breakage) and macroscopic behavior (displacement weakening).} 
\end{figure}

The quantity $x_p (v)$ marks the transition from a strengthening to a weakening behavior, associated with the gradual detachment of polymer chains. To check this assertion, the solid curves in Fig.~\ref{fig:nxplot} show the evolution of the fraction of initially intact glue strands that remain attached to the collagen fibrils at both ends (i.e. with end bonds being intact), as a function of the displacement $x$, at the same pulling velocity $v = 1000$ nm s$^{-1}$. One sees immediately that the onset of polymer chain detachment coincides with the transition to weakening behavior at $x_p$. In addition, under our assumption of a uniform distribution of polymer chain lengths, the number of polymer chains that have yet to rupture decreases linearly with separation distance in the weakening regime. On the other hand, the dashed curves show that breakage of sacrificial bonds within individual polymer chains occurs continually in both the strengthening and weakening regimes. Figure~\ref{fig:nxplot} thus demonstrates how microscopic physics (bond breakage), characterized by the state variables $N_b$ and $N_e$, accounts for macroscopic behavior (displacement strengthening and subsequent weakening) in biological structures that contain sacrificial bonds and hidden length. In the context of dynamic fracture, the strengthening regime corresponds to crack arrest and the weakening regime corresponds to crack propagation and catastrophic failure.

\begin{figure}[here]
\centering 
\subfigure{\includegraphics[width=.45\textwidth]{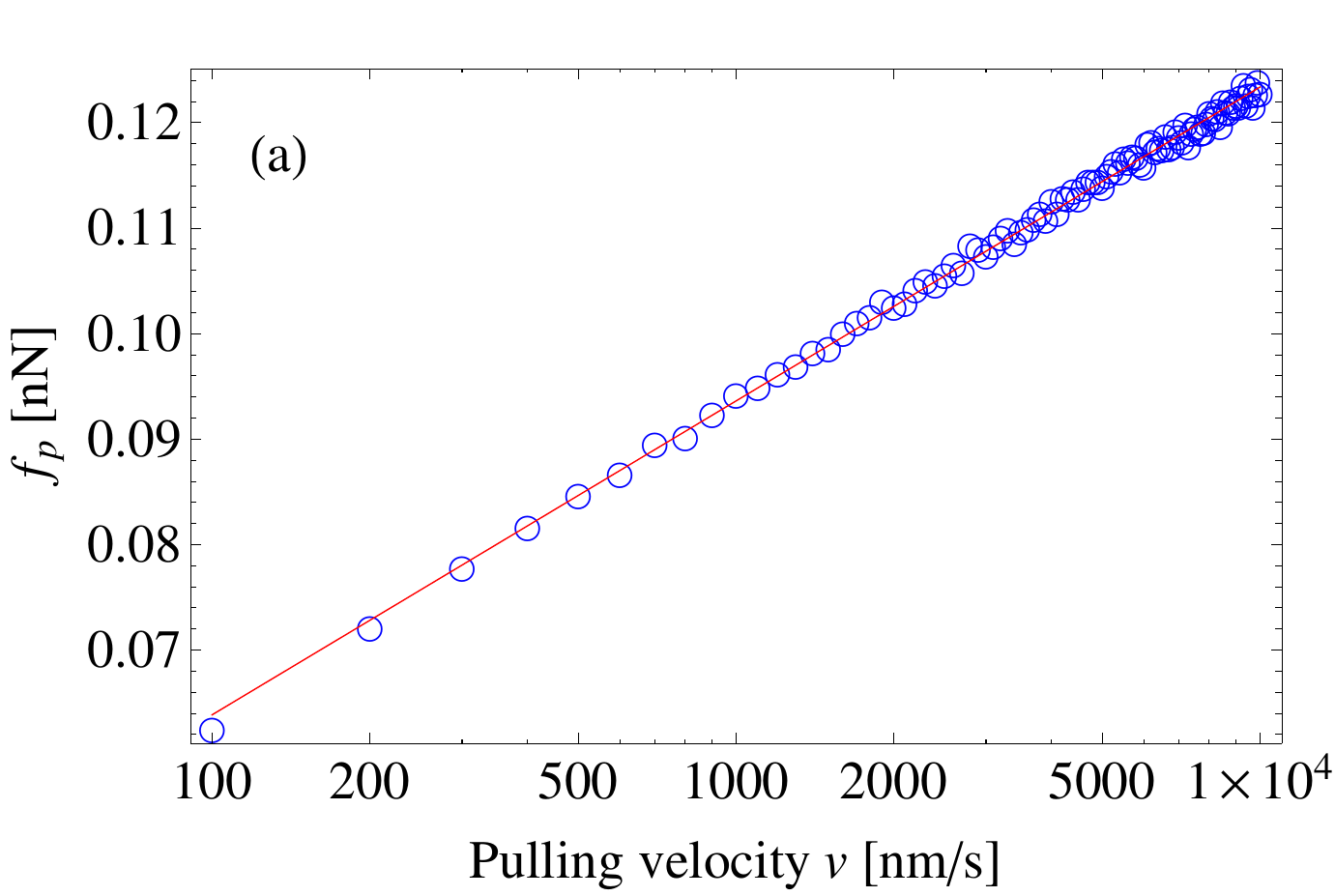} \label{fig:fvfit}}
\subfigure{\includegraphics[width=.45\textwidth]{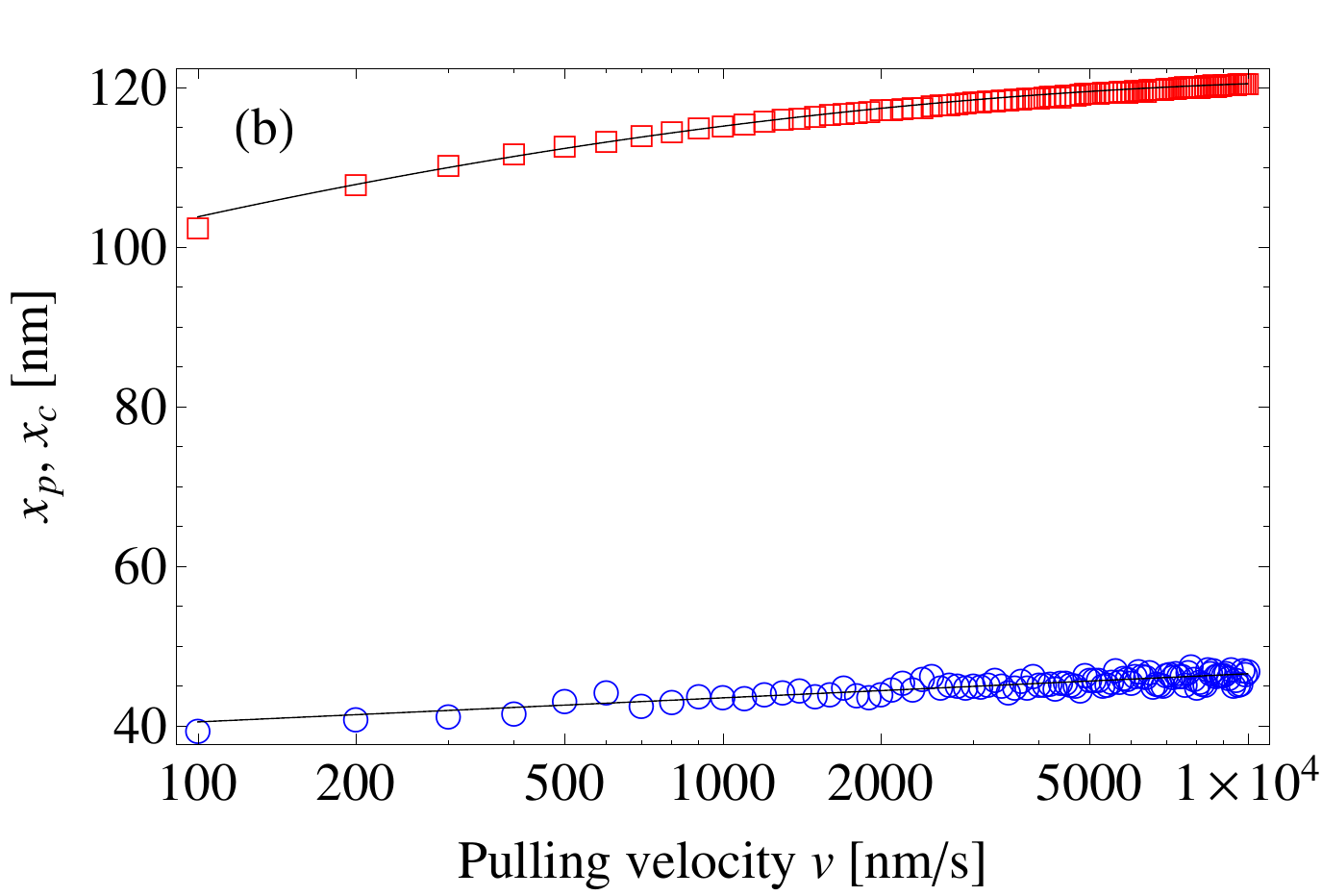} \label{fig:xvfit}}
\caption{(Color online) (a) Plot of the average peak force $f_p (v)$ per polymer chain as defined in Eq.~\eqref{eq:fxv} versus the pulling speed $v$. The blue data points are average values from 100 runs and the red curve is the log-linear fit given by Eq.~\eqref{eq:fp}. (b) Plots of the displacement at peak force $x_p (v)$ (blue circles) and maximum pulling distance $x_c (v)$ (red squares) as defined in Eq.~\eqref{eq:fxv} versus the pulling speed $v$. The data points are average values from 100 runs. The black curves are the fits given by Eqs.~\eqref{eq:xp} and~\eqref{eq:xc}.} 
\end{figure}

To extract the velocity dependence of the quantities $f_p (v)$, $x_p (v)$ and $x_c (v)$ explicitly, we plot these quantities in Figs.~\ref{fig:fvfit} and~\ref{fig:xvfit}. They can be fit with the functional forms
\begin{eqnarray}
 \label{eq:fp} f_p(v) &=& f_1 \log \left( \dfrac{v}{v_0} \right) + f_0 ; \\
 \label{eq:xp} x_p(v) &=& p_1 \log \left( \dfrac{v}{v_0} \right) + p_0 ; \\
 \label{eq:xc} x_c(v) &=& c_2 \left[ \log \left( \dfrac{v}{v_0} \right)^2 \right] + c_1 \log \left( \dfrac{v}{v_0} \right) + c_0.
\end{eqnarray}
Within our choice of system parameters, we find $v_0 = 100$ nm s$^{-1}$, $f_0 = 0.064$ nN, $f_1 = 0.013$ nN, $p_0 = 41$ nm, $p_1 = 1.3$ nm, $c_0 = 104$ nm, $c_1 = 6.3$ nm and $c_2 = -0.57$ nm.

We have thus shown that in the limit of long delay times $t$, the total force $F$ on the ensemble of polymers can be factored into the product of the number $N_p N_e^* (t)$ of polymer chains that are intact at the beginning of the pulling experiment, times the average force $f(x, v)$ per polymer chain. Both the strengthening and weakening regimes -- the latter being associated with the rupture of polymer chains and their detachment from the substrate -- can be described by power laws characterized by the velocity-dependent peak force $f_p (v)$, the displacement at peak force $x_p (v)$ and the maximum pulling distance $x_c (v)$. Among these, the peak force $f_p$ varies linearly with the logarithm of the pulling velocity $v$, in conformity with experiments. This constitutive approach enables us to describe the mean-field dynamics of sacrificial bond breakage and hidden length release using several adjustable parameters, without having to account for the random breakage of individual bonds in detail.

\section{Concluding Remarks}

In this paper we have developed a simple quasi-one-dimensional kinetic model, based on Bell's theory, that describes the breakage of sacrificial bonds and release of hidden length in biological structures such as the linkage between collagen fibrils in animal bone. The kinetic model draws ideas from theories of protein unfolding, a process that also involves the forced rupture of noncovalent bonds and the exposure of folded structure. It tracks the evolution of the number of sacrificial bonds $N_b$ and the number of chains $N_e$ that adhere to the substrate -- the only two molecular-state-dependent variables in the theory -- with the pulling distance $x$ and the total force $F$ on the polymer network, according to a velocity-dependent criterion that determines the times or displacements at which bond breakage occurs. The force is entropic in nature, given by the wormlike-chain model as a function of the pulling distance $x$ and the amount of available contour length $L$, the latter of which is computed in terms of the number of remaining sacrificial bonds $N_b$.

We have shown that sacrificial bonds and hidden length lead to a marked increase in fracture toughness in materials where they are present, and that both the fracture toughness and maximum displacement before complete rupture in a pulling experiment increases with the pulling velocity $v$ which drives the system away from equilibrium. In particular, the peak force $f_p$ in the force-displacement profile varies linearly with the logarithm of the pulling velocity $v$, in conformity with various mechanical experiments on biological molecules such as~\cite{adams_2008}. In addition, our kinetic model naturally incorporates self-healing, evidenced by the increase in the number of attached polymer chains, rupture peak height and total fracture toughness with recovery time. Our simple quasi-one-dimensional model, however, does not explicitly account for the effect of crosslinks and entanglements in a network of glue strands. The extent to which these additional microscopic mechanisms will impact the macroscopic behavior will be investigated in future work.

Based on our theoretical calculations we have proposed a phenomenological description for the force-displacement profile of a collection of polymer chains with a distribution of lengths. The force-displacement profile consists of a strengthening regime for small displacements, where the force increases with the displacement according to a power law. This is followed by a weakening regime associated with the gradual detachment of polymer chains which no longer contribute to force transmission. Such a constitutive description will be of utmost utility in future multiscale simulations of bone fracture. The dynamical behavior of glue connection between collagen fibrils has important implications on crack propagation, crack arrest, strength recuperation and collagenous diseases in bone.

\section*{Acknowledgments}

We thank Paul Hansma, Georg Fantner, James Langer and Megan Valentine for instructive discussions and comments. This work was supported by the David and Lucile Packard Foundation, a UC Santa Barbara Dean's Fellowship, Office of Naval Research MURI grants N000140810747, and the Institute for Collaborative Biotechnologies through contract no. W911NF-09-0001 from the U.~S.~Army Research Office. The content of the information does not necessarily reflect the position or the policy of the Government, and no official endorsement should be inferred.

\end{document}